\documentclass{epl}

\usepackage{psfrag}
\usepackage{amssymb}

\title{First- and Second Order Phase Transitions in the
  Holstein-Hubbard Model}
\shorttitle{Hubbard-Holstein Model}
\author{W. Koller\inst{1} \and D. Meyer\inst{1} \and Y. \=Ono \inst{2}
  \and A. C. Hewson \inst{1} }
\institute{
  \inst{1} Department of Mathematics, Imperial College, London SW7 2BZ, UK\\
  \inst{2} Department of Physics, Niigata University, Ikarashi, Niigata 950-2181, Japan
}

\begin{document}

\maketitle

\begin{abstract}
We investigate metal-insulator transitions in the half-filled Holstein-Hubbard
model as a function of the on-site electron-electron interaction $U$
and the electron-phonon coupling $g$.
We use several different numerical methods to calculate the phase
diagram, the results of which are in excellent agreement.
When the electron-electron interaction $U$ is dominant the transition
is to a Mott-insulator; when the electron-phonon interaction
dominates, the transition is to a localised bipolaronic state.
In the former case, the transition is always found to be second order.
This is in contrast to the transition to the bipolaronic state, which
is clearly first order for larger values of $U$.
We also present results for the quasiparticle weight and the
double-occupancy as function of $U$ and $g$.
\end{abstract}

Strong correlation effects and localization can occur in metallic systems
due both to strong electron-electron interactions and strong
electron-phonon coupling and also their interplay.
There are many systems with strongly correlated electrons where there
is also a strong coupling to the lattice and lattice modes, for
example \chem{V_2O_3}\cite{MK94,IFT98}, manganites\cite{MLS95} or
fullerides\cite{Gun97}.
The strong electron-electron interactions can be described by the Hubbard model,
where the transition to a Mott-Hubbard insulator  has been extensively
investigated\cite{Hub64b,Jar92,GKKR96,MSKRF95,Bul99}.
The Holstein model has been used to examine localization to a polaronic or
bipolaronic insulator due to electron-phonon interactions\cite{Hol59,BZ98,MHB02,CC03}.
The interplay between these two types of localization can be investigated
using the Holstein-Hubbard model which includes both types of interaction:
\begin{equation}
  \label{eq:hamil}
  H=\sum_{\vec{k}\sigma} \epsilon(\vec{k}) c_{\vec{k}\sigma}^\dagger
  c_{\vec{k}\sigma} +
        U \sum_{i} n_{i\uparrow} n_{i\downarrow} +
 \omega_0 \sum_i  b_i^\dagger b_i + g\sum_i  (b_i^\dagger +b_i) \big(\sum_{\sigma}
  n_{i\sigma} -1 \big),
\end{equation}
where $U$ describes the electron-electron interaction within a semielliptic band of
dispersion $\epsilon(\vec{k})$ and band width $W=4$, $g$ the electron-phonon
coupling and $\omega_0=0.2$ is the frequency of a local Einstein-phonon mode.
In the large-$\omega_0$ limit $(\omega_0 \gg W)$, the model can be mapped onto an effective
Hubbard model with $U_{\rm eff} \equiv U-2 g^{2} / \omega_{0}$\cite{Fre93,MH03}.

\begin{figure}
\onefigure[scale=0.8]{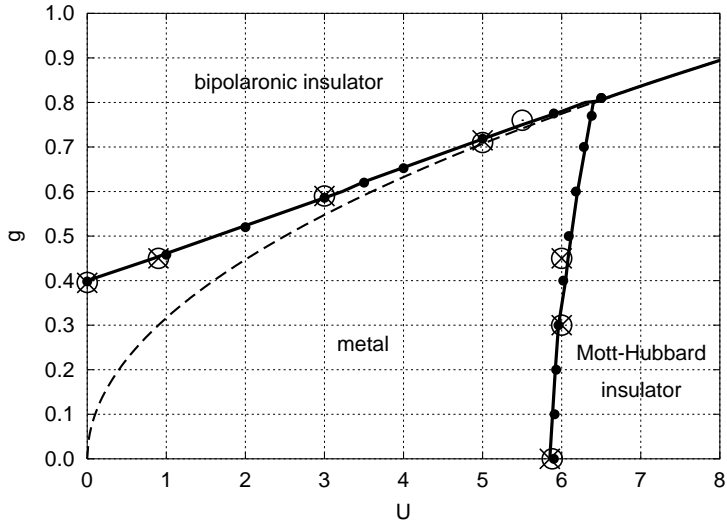}
\caption{The calculated phase diagram of the Hubbard-Holstein model at
 $T=0$. The solid lines represent the DIA2 results. They are in excellent
 agreement with the DIA4, NRG, and ED results, as indicated by the crosses,
 open circles, and filled circles respectively.
 The dashed line represents the locus of the points with
 $U_{\rm eff} \equiv U-2 g^{2} / \omega_{0} =0$ which becomes the
 phase boundary for $U>6.4$.}
\label{phasedia}
\end{figure}

Here we investigate the phase diagram in the particle-hole symmetric
case. We neglect phases of long-range order and concentrate on metal
to insulator transitions driven by both electron-electron and
electron-phonon interactions\cite{JPHLC03pre}. 
In general, no exact solution of this model is known. We use a number of
approximation schemes which lead to a consistent picture of the physics of
these transitions.
Two of these approaches are based on the dynamical mean-field theory
\cite{GKKR96}, where the lattice model is mapped onto an effective impurity
model with a self-consistency constraint.
We apply two different methods to solve the effective impurity model, the
exact diagonalization method (ED) \cite{CK94} and
the numerical renormalization group (NRG) \cite{KWW80a,BHP98}.
In the ED method a restricted basis set is used to describe the bath of the
impurity model, which can then be solved exactly.
The ED calculations presented here were performed on an $8$-site cluster.
In the NRG approach, the impurity model is solved by an iterative diagonalization
scheme which has the advantage that it can probe arbitrarily low energy
scales.
In the NRG calculations up to 1200 states were retained with $\Lambda=1.8$ and
up to $36$ phonon states.
We also use the dynamical impurity approximation (DIA),
introduced recently by Potthoff \cite{Pot03,Pot03b}.
This method has the advantage of being a variational approach for the grand
potential. Its main limitation is the restricted set of trial selfenergies
that one can handle in practice. We use trial selfenergies
corresponding to impurity models with one (two-site DIA -- DIA2) and three
bath sites (four-site DIA -- DIA4).

In figure \ref{phasedia}, we plot the $g$ vs. $U$ phase diagram as
obtained by the methods described above. 
There is excellent agreement between the results of all four methods.
The critical coupling for the Mott transition $U_{cM}\approx 5.85$ on the $g=0$
axis corresponds to the results already known for the Hubbard
model\cite{MSKRF95,Bul99}, and $g_{c}\approx0.39$ on the $U=0$ axis for the
Holstein model \cite{MHB02,CC03}.
Within the DMFT for $U<U_{cM}$ there is a range of values of $U$ where the
insulating solution coexists with the metallic one (''hysteresis''), but for
$T=0$ the metallic solution is always the physically relevant one.
For the Holstein model a significant coexistence region exists only for large
$\omega_0$\cite{MH03}.
The dashed line (polaronic line) is the locus of the points with $U_{\rm eff}=0$;
above (below) this line, $U_{\rm eff} < 0$ ($> 0$).  
The metal-insulator transition in the region $U_{\rm eff} >0$ is very similar
to the Mott-Hubbard transition as found in the pure Hubbard model, where
double-occupancy $d=\langle n_{i\uparrow}n_{i\downarrow}\rangle$ is almost
completely suppressed.

The phase boundary to the Mott-Hubbard insulator is largely independent of the 
electron-phonon coupling until the polaronic line is reached. This is
due to the fact that in this
region the charge fluctuations which couple to the phonons are
strongly suppressed.
On the other
hand, the transition in the other region ($U_{\rm eff} < 0$) depends strongly
on both $U$ and $g$. The insulating phase here is a bipolaronic one with
enhanced on-site double-occupancy.

\begin{figure}
\twoimages[scale=0.52]{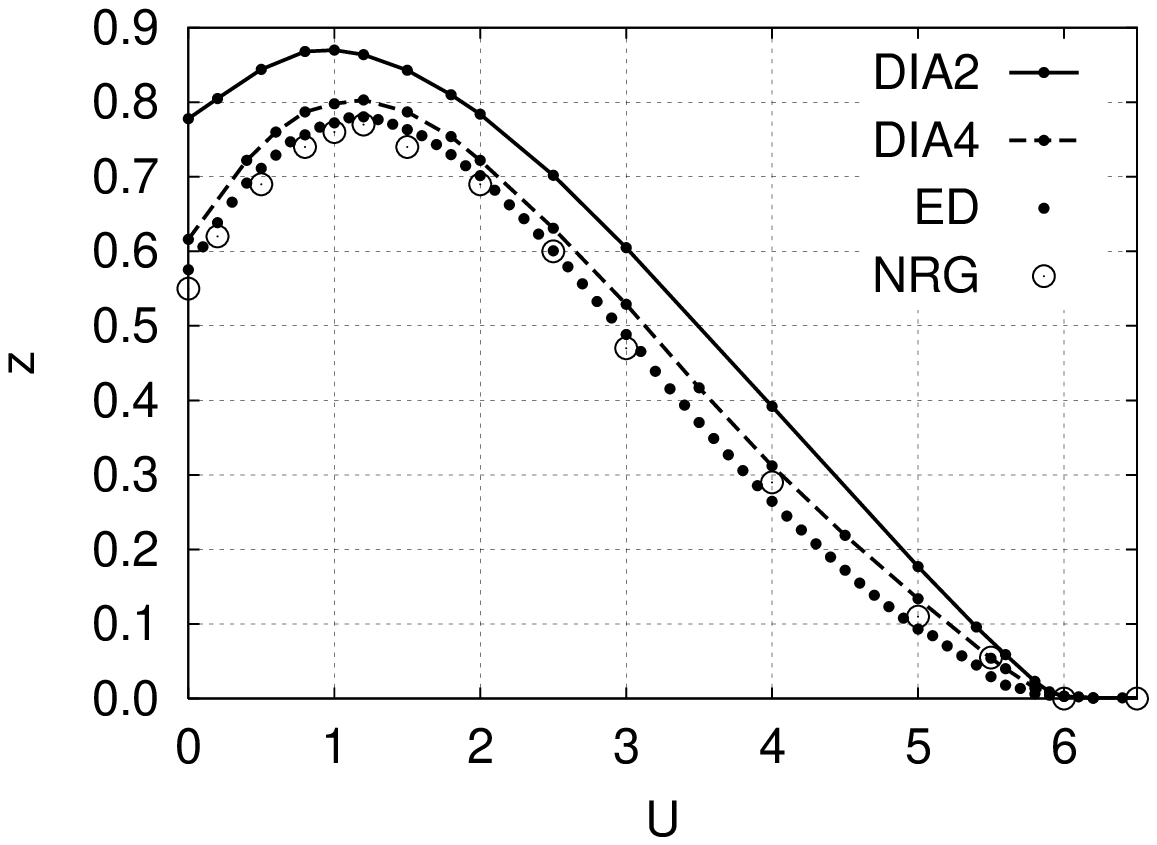}{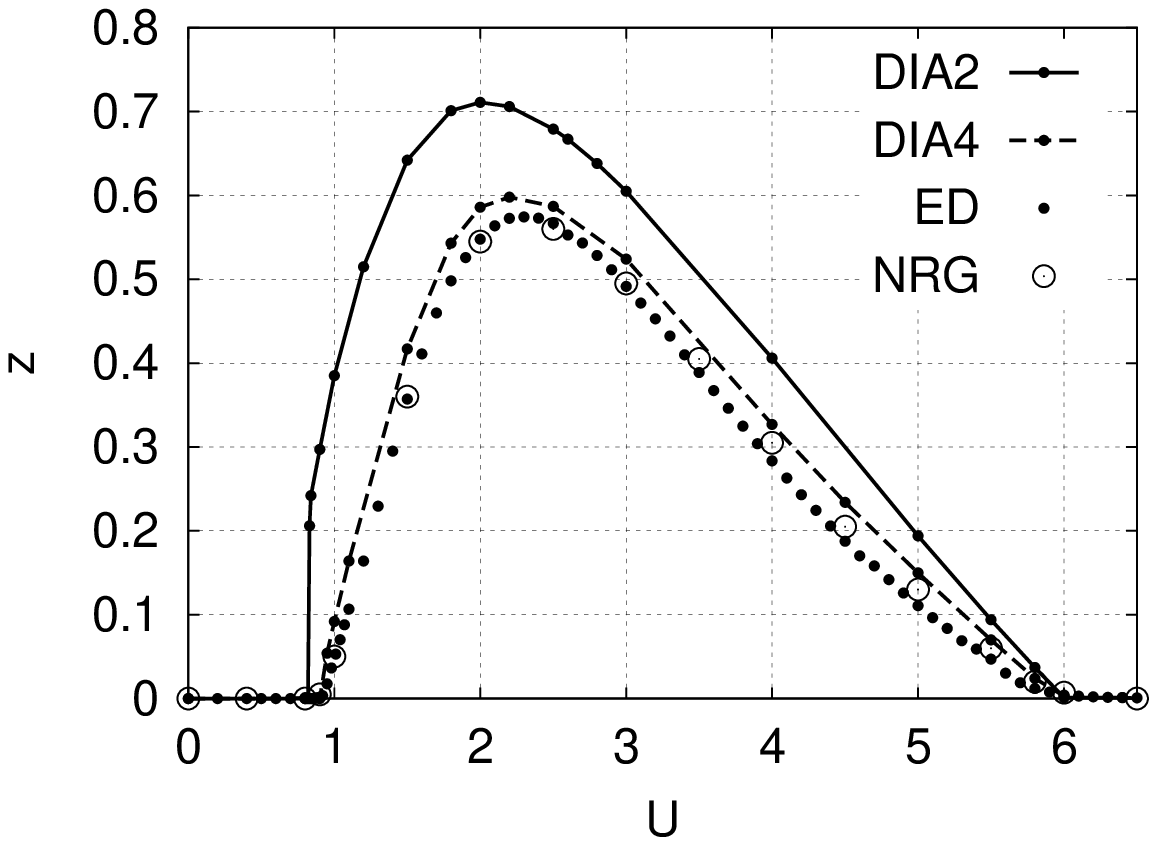}
\caption{The quasiparticle weight $z$ as a function of the electron-electron
  repulsion $U$ for two values of the electron-phonon coupling:
  $g=0.3$ in the left figure and $g=0.45$ in the right.}
\label{zU}
\end{figure}

To probe the physics of these transitions in more detail, we consider the
variation of the quasiparticle weight $z$ both in scans with a fixed $g$ or a
fixed $U$. In figure \ref{zU} we look at the cases $g=0.3$ and $g=0.45$.
The results of all four methods show the same general trend, although the
two-site DIA appears to systematically overestimate the values of $z$ in
general but gives almost correct values for the critical couplings.
For $g=0.3$ we start in a metallic state for $U=0$ with a value of
$z$ which is already rather less than one, due to the electron-phonon interaction
\cite{MHB02}. Upon increasing $U$, $z$ initially increases until $U_{\rm
  eff}\approx 0$ and then steadily decreases to $z=0$ at the critical value of
$U=U_{cM}$. 
For $g=0.45$ and $U=0$ we start in the insulating phase. At a critical value
$U=U_{cB}\approx 0.9$ there is a sudden onset of a metallic solution which
suggests that this transition might be first order. As in the case for
$g=0.3$, the quasiparticle weight increases until $U_{\rm eff}\approx 0$ and
decreases to the transition at $U=U_{cM}\approx 6.0$. In both cases, $z$
vanishes continuously as $U$ approaches $U_{cM}$ from below.

\begin{figure}
\twoimages[scale=0.52]{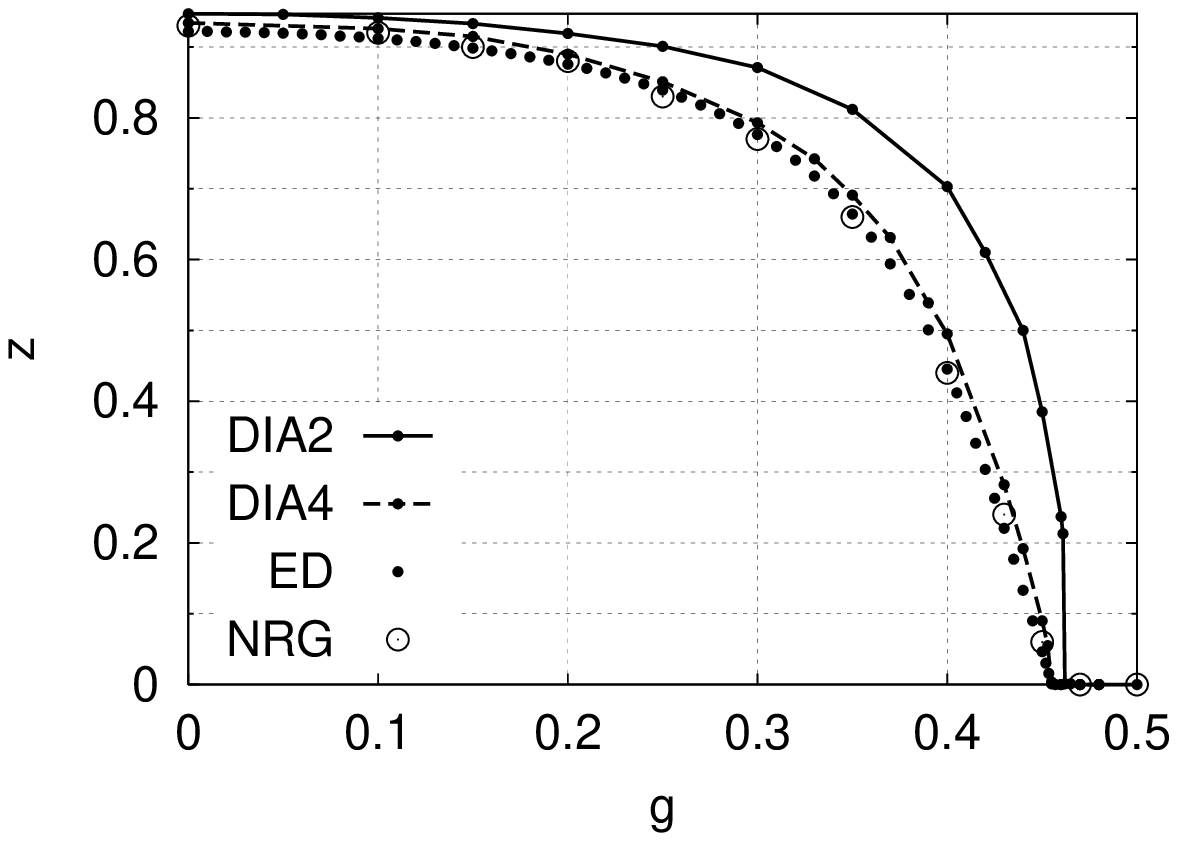}{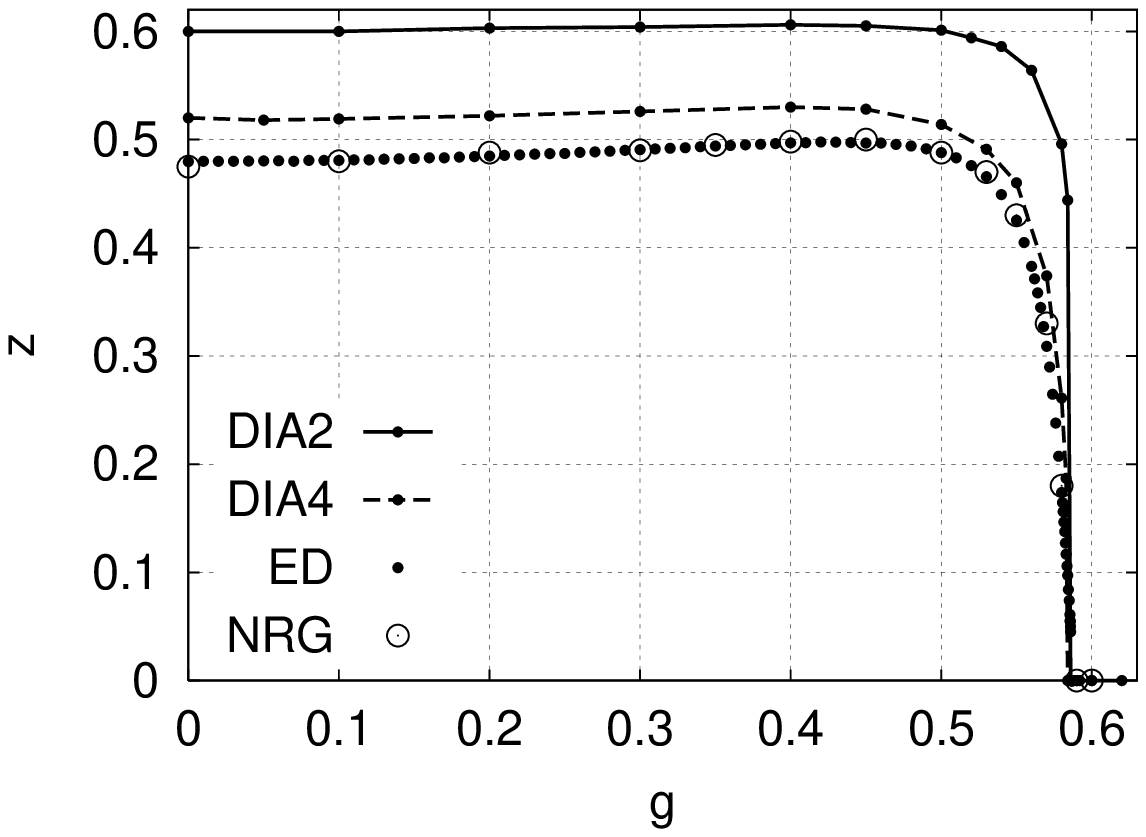}
\twoimages[scale=0.52]{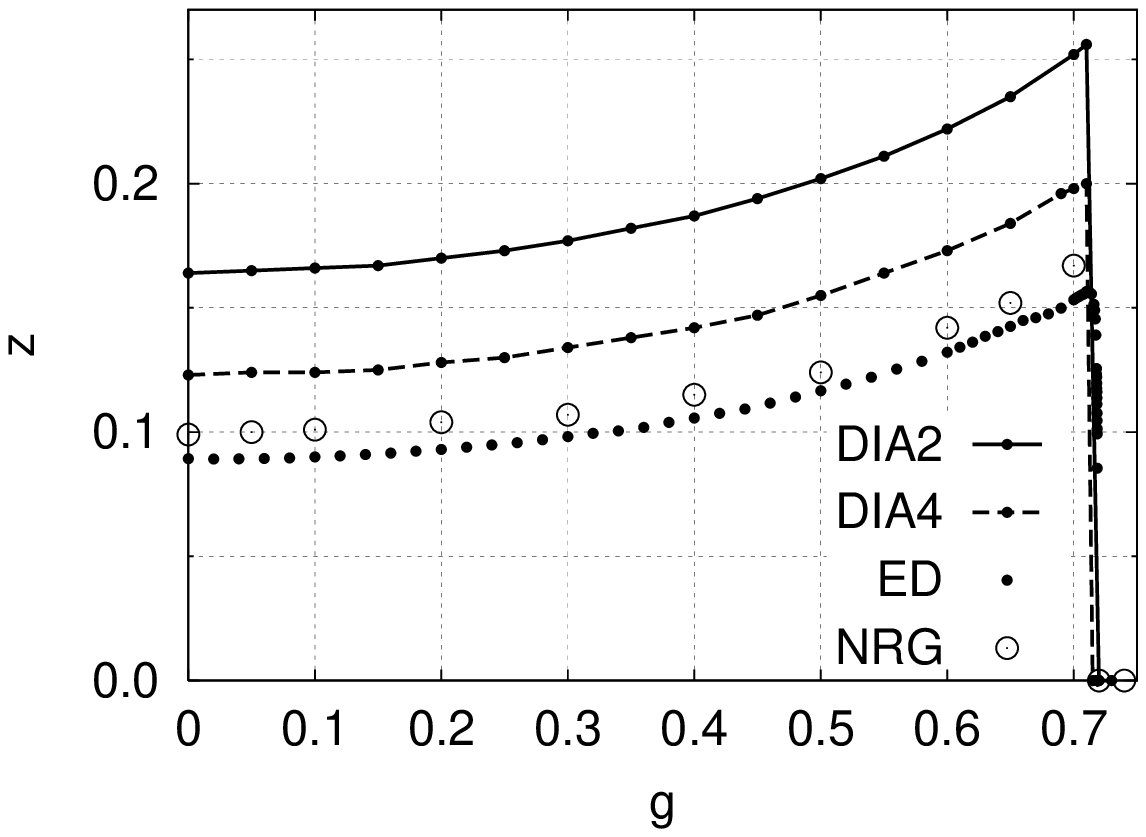}{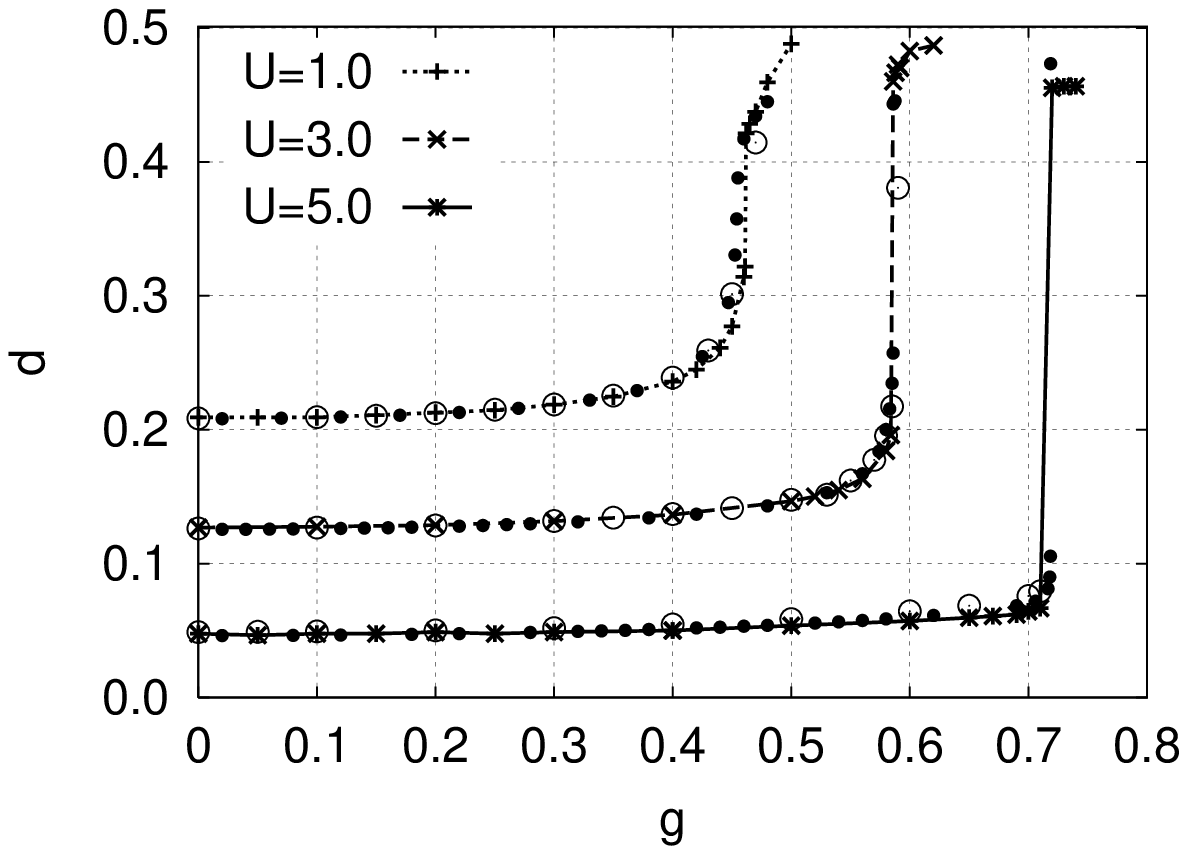}
\caption{The quasiparticle weight $z$ as a function of the electron-phonon
  coupling $g$ for three values of the electron-electron interaction:
  $U=1.0$ in the upper left, $U=3.0$ in the upper right and $U=5.0$ in
  the lower left figure. The lower right figure shows the
  double-occupancy $d=\langle n_{\uparrow} n_{\downarrow} \rangle$ as
  a function of $g$ for all three values of $U$. The lines correspond
  to the DIA2 results and the open (filled) circles to the NRG (ED) results.} 
\label{zg}
\end{figure}

The complementary scans of $z$ with fixed $U$ and variable $g$ are shown in
figure \ref{zg}.
There is little change of $z$ in the region with $U_{\rm eff} >0$ as the
repulsive electron-electron interaction $U$ effectively suppresses charge
fluctuations which would otherwise couple to the phonons.
This suppression can also be seen in the results for the
double-occupancy $d=\langle n_{\uparrow}n_{\downarrow}\rangle$, which
are plotted in figure \ref{zg}. 
Once $U_{\rm eff}$ is less than zero, there is a rather rapid decrease
in $z$ and an apparent jump at the critical value of $g$ indicating a
first order transition. There is a corresponding sudden increase of
the double-occupancy $d$. For larger values of $U$, $z$ even increases
until a point very close to the transition, and the jump becomes more
pronounced as can be seen for $U=5.0$ (lower left figure \ref{zg}). 
Above the transition the value of $d \approx 0.5$.

\begin{figure}
\psfrag{Omega}{\hspace*{-1.2em}$\Omega-\Omega(V\!\!=\!0)$}
\psfrag{V}{$V$}
\twoimages[scale=0.52]{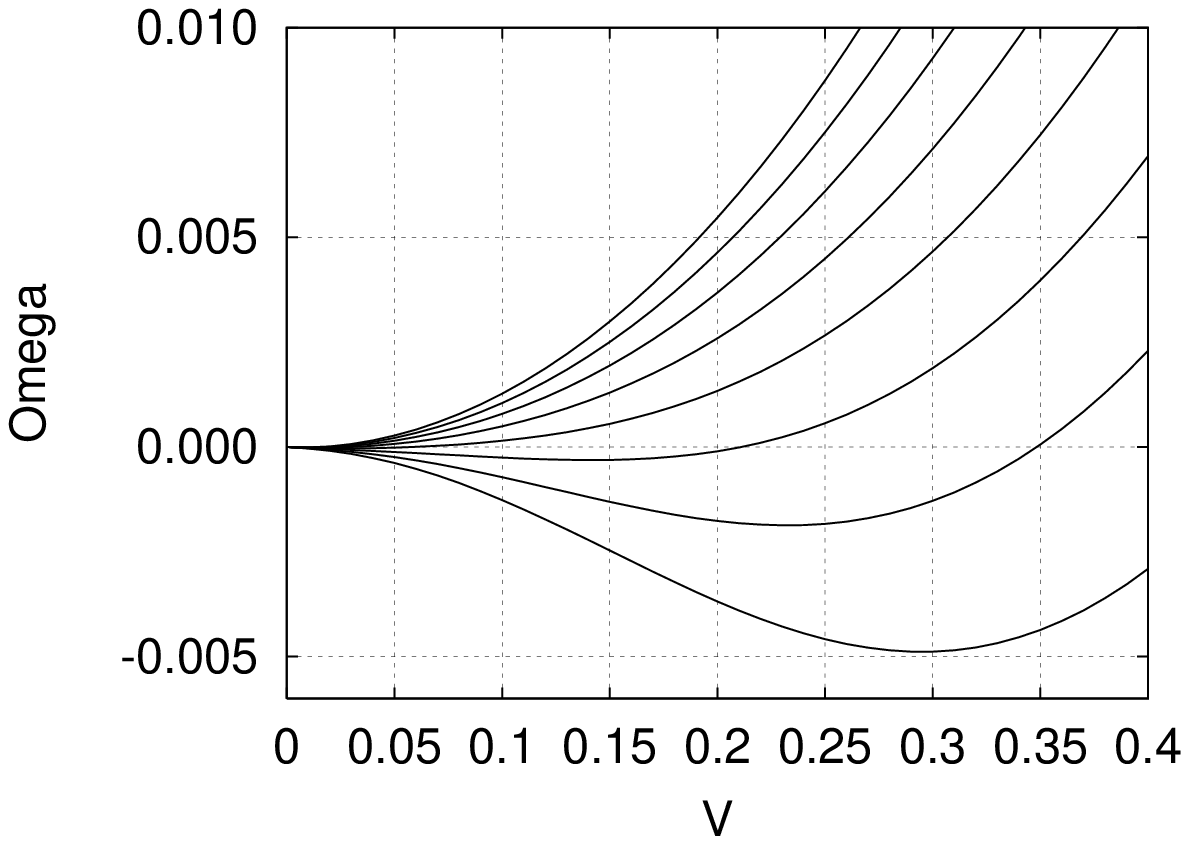}{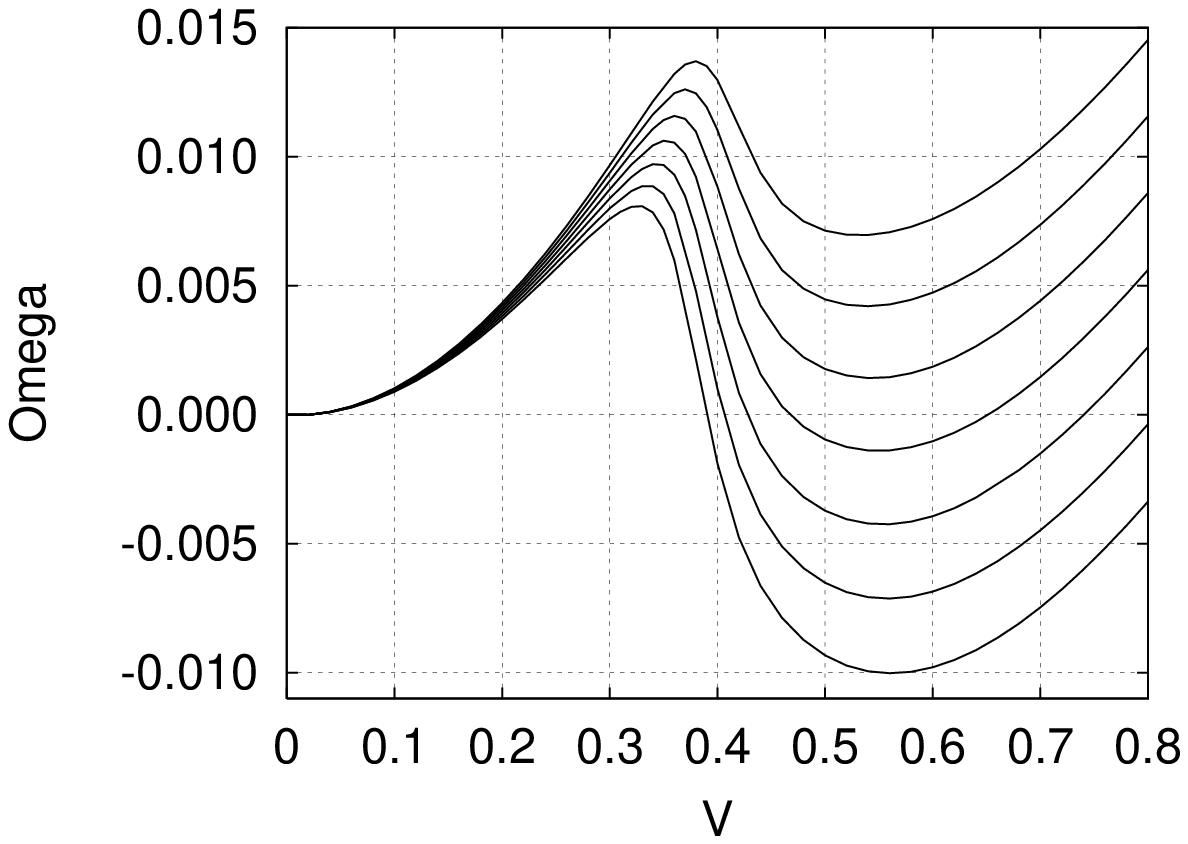}
\caption{The grand potential $\Omega$ as a function of the variational
  parameter $V$ of the DIA2 method at $g=0.60$.
  The left figure displays a second order transition from a metal to a
  Mott-Hubbard insulator.
  The minimum shifts smoothly to zero as $U$ increases from $U=5.6$,
  the lowest curve, in steps of $0.2$ to
  $U=7.0$ for the uppermost curve.
  The right figure shows a first order transition from a metal to a
  bipolaronic state.
  There is a jump of the position of the minimum as $U$ decreases from
  $U=3.26$, the lowest curve, in steps of $-0.01$ to $U=3.20$ for
  the uppermost curve.}
\label{omega}
\end{figure}

With numerical methods it is difficult to exclude the possibility that the
transition is continuous but very sharp.
However, with the DIA one calculates the grand potential
$\Omega$ which enables one to address the order of the transition directly.
In figure \ref{omega} we plot the DIA2 results for $\Omega$ as a
function of the variation parameter $V$ for $g=0.60$ and values of $U$
close to the transitions at $U_{cM}$ (left plot) and $U_{cB}$ (right plot).
For the transition at $U_{cM}$ we see that there is a global minimum which
shifts continuously to $V=0$ as $U$ approaches $U_{cM}$. In the DIA2, $V=0$
corresponds to the insulating solution. These results are very similar to the
results obtained by Potthoff for the pure Hubbard model \cite{Pot03b}. 

In the right-hand figure the same data is plotted for the transition at
$U_{cB}$. Starting with the largest value of $U=3.26$       
the global minimum is at finite $V\approx 0.56$             
corresponding to a metallic state. Decreasing $U$, the minimum
becomes shallower but shifts only slightly to smaller values $V\approx0.54$.  
The minimum at $V=0$ becomes the global minimum for $U<U_{cB}=3.225$.
This is clear evidence for a first order transition, at least within the DIA2.
Coexistence of metallic and insulating solutions can be found there, as is
corroborated by the other methods.

\begin{figure}
\psfrag{Omega}{\hspace*{-1.6em}$\Omega-\Omega_{2}(V\!\!=\!0)$}
\psfrag{g}{$g$}
\onefigure[scale=0.8]{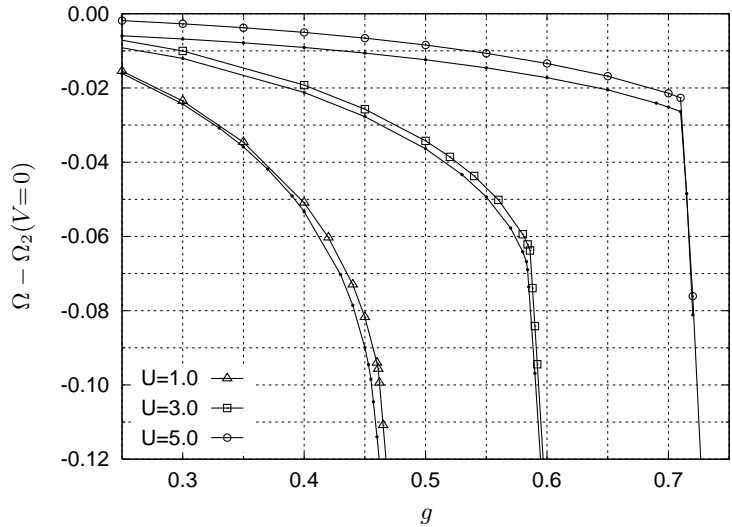}
\caption{The grand potential $\Omega$ as function of $g$ for three
  different values of $U$, as calculated with the DIA2 (big open symbols)
  and DIA4 (small dots) methods.
  Here, $\Omega_2(V=0)$ denotes the value of the grand potential for
  $V=0$ in the DIA2 method.}
\label{omg}
\end{figure}

In figure \ref{omg} we plot the grand potential as a function of $g$
as calculated with the DIA2 and DIA4 approaches for three values of
$U=1.0, 3.0$ and $5.0$.
For the largest value of $U=5.0$, the  discontinuity of the gradient
of $\Omega$ with respect to $U$ is clearly seen.
As one reduces $U$, the kink becomes progressively smaller. 
For DIA2 calculations, similar to those presented in figure
\ref{omega}, a kink persists down to $U=0$ and the phase transition is always
first order. 
For the corresponding DIA4 calculations, however, the situation is not
quite so clear.
In this small-$U$ range, a jump in $z$ is also difficult to identify (see
figure \ref{zg}). The evidence from the calculations apart from the
DIA2, indicates that the transition is probably for all practical
purposes continuous for $U < 3$. 
In contrast, the transitions along the line $U_{cM}$ are always continuous as
a function of $U$.

All methods used in our investigation lead to a consistent picture of the
metal-insulator transitions in the Holstein-Hubbard model. 
The phase diagram is composed of three regions, a metallic region and two
qualitatively different insulating regions. The boundary between the metallic
and the Mott-insulating region is described by a line of continuous
transitions, and is only weakly dependent on the electron-phonon
coupling.
Our evidence is that the transition from the metal to a bipolaron
insulator state, which is characterised by an enhanced on-site
double-occupancy, is of first order for larger values of $U$.
However, the discontinuity becomes progressively smaller as $U$ is
decreased and has virtually disappeared for the range $U < 3$.

The behaviour near the boundary between the two insulating phases is difficult
to access by all of our methods except within the DIA2, where the results
indicate that there is a first order transition between them.
Detailed results and discussions of dynamical response functions at $T=0$ for
this model have been calculated and will be prepared for publication in due
course. 

\acknowledgments
We wish to thank the EPSRC (Grant GR/S18571/01) for financial
support. One of us (Y\=O) was supported by the 
Grant-in-Aid for Scientific Research from the Ministry of Education, 
Culture, Sports, Science and Technology.
We also thank M. Aichhorn, R. Bulla and  M. Potthoff for helpful discussions.

\bibliographystyle{../../Lit/uuu.pott}
\bibliography{../../Lit/Own_article.bib,../../Lit/artikel.bib,../../Lit/artikel.nicht.vorhanden.bib,../../Lit/buecher.bib,../../Lit/arbeiten.bib}

\end{document}